\documentclass[12pt,letterpaper]{article}

\usepackage{amsmath,amssymb,array,calc,rotating,epsfig,psfrag, amscd}
\usepackage{color}
\usepackage{hyperref}
\usepackage[left=2cm,top=2cm,right=3cm,nohead]{geometry}

\newcommand{\bea}{\begin{eqnarray}}
\newcommand{\eea}{\end{eqnarray}}
\newcommand{\be}{\begin{equation}}
\newcommand{\ee}{\end{equation}}
\newcommand{\barr}{\begin{array}}
\newcommand{\earr}{\end{array}}

\newcommand{\bpm}{\begin{pmatrix}}
\newcommand{\epm}{\end{pmatrix}}
 \newcommand{\bitem}{\begin{itemize}}
 \newcommand{\eitem}{\end{itemize}}

\definecolor{cardinal}{rgb}{0.6,0,0}
\definecolor{darkgreen}{rgb}{0,0.5,0}
\definecolor{golden}{rgb}{0.92, 0.7, 0}
\definecolor{midnight}{rgb}{0, 0, 0.5}
\definecolor{darkblue}{rgb}{0.2, 0, 0.8}

\newcommand{\beq}{\begin{equation}\begin{aligned}}
\newcommand{\eeq}{\end{aligned}\end{equation}}

\begin{document}

\begin{flushright}
YITP--SB--12--15
 \end{flushright}

\vspace{0.5cm}
\begin{center}

{\Large \bf A Universal Formula for the Stress--Tensor\\
\vskip 3mm
Contribution to Scalar Four--Point Functions}\\
\vskip 3mm

\vskip1.7cm 
{\bf Gregory Giecold$^{*}$}
 
 \vskip0.7cm
\textit{C.N.~Yang Institute for Theoretical Physics,\\
State University of New York,\\
Stony Brook, NY 11794--3840\\ U.S.A.}\\

\let\thefootnote\relax\footnotetext{

  $^{*}$\href{mailto:giecold@insti.physics.sunysb.edu}{{\tt giecold@insti.physics.sunysb.edu}}}

\end{center}

\vskip1.7cm
\begin{abstract}
We illustrate the power and efficiency of a recently uncovered Mellin--space approach to AdS/CFT correlation functions by providing a universal formula for the 4--scalar graviton--exchange Witten diagram, for arbitrary CFT--dual scaling dimensions. Our result keeps the space--time dimension generic as well, and is expressed as a combination of just 11 hypergeometric functions. Such hypergeometric functions are related to scalar--exchange diagrams. In particular, if reverse--engineered in terms of D--functions, this might be viewed as a first--step towards proving a long--standing conjecture by Dolan, Nirschl and Osborn pertaining to four--point correlators of chiral primary operators at strong coupling. Most importantly, the technology developed herein marks an additional development towards the long--anticipated computation of the four--point function of the $\mathcal{N}=4$ sYM stress--tensor.
\end{abstract}

\newpage

\section{Introduction}

According to the gauge/gravity duality, supergravity scattering amplitudes in a $d+1$--dimensional $\text{AdS}_{d+1}$ background encode the planar contribution to correlation functions of operators of a dual conformal field theory $CFT_{d}$ at strong coupling. Yet, despite more than a decade of efforts, the evaluation of AdS/CFT correlation functions was until very recently a difficult and computationally intensive task. 

Such calculations were typically done in position space. It is natural to wonder if this is the right language. In the same way that going to momentum space simplifies the evaluation of Feynman diagrams in flat space, one might suspect that, similarly, different variables could possibly be lurking around, which would facilitate the computation of Witten diagrams. 

Indeed, it has recently been illustrated through various examples that applying a Mellin--transform to AdS/CFT correlators dispenses with a lot of hard work and sheds light on their structure~\cite{Mack:2009mi, Mack:2009gy, Penedones:2010ue, Paulos:2011ie, Fitzpatrick:2011ia}. In particular, the contents of operator product expansions is disentangled after going to Mellin space. This boils down to the fact that Mellin amplitudes are meromorphic with simple poles whose positions correspond to the scaling dimensions of operators in the OPE~\cite{Fitzpatrick:2011hu, Fitzpatrick:2011dm}. In a sense, going to Mellin space realizes a natural basis of OPE, as first envisioned by Polyakov~\cite{Polyakov:1974gs}.

So far, the Mellin--space approach has mostly focused on evaluating Witten diagrams involving scalar fields only. Feynman--like rules for tree--level Mellin amplitudes have been proven~\cite{Nandan:2011wc} in this case. The purpose of this short note is to show that Mellin--space is just as well apposite to handling with great efficiency processes involving the exchange of a graviton in the bulk. We managed to find a universal formula that captures at one fell swoop the stress--tensor exchange contribution to any 4--point function of scalar operators, no matter the particular values of their conformal weights or of the space--time they live in. This should be contrasted with the great amount of effort that is exacted by a position--space approach to evaluating such diagrams, not to mention that this modus operandi cannot achieve the universality of our formula~\cite{D'Hoker:1999pj, Arutyunov:2000py, Arutyunov:2002fh, Arutyunov:2003ae, Berdichevsky:2007xd, Uruchurtu:2008kp, Uruchurtu:2011wh}. 

It bears noting that the validity of our formula is rigorously established by setting all the operator dimensions to $\Delta_i = \Delta = d$ in $d=4$ dimensions, and comparing with equation (8) from~\cite{Penedones:2010ue}, which is equation~\eqref{mincoupled} of the present publication. 

Three--point functions of chiral primary operators are determined by superconformal invariance. Therefore, their supergravity computation does not tell much\footnote{As expected, this agrees with a free--field computation~\cite{Lee:1998bxa}.}, beyond providing further evidence for the validity of the AdS/CFT correspondence, which, of course, is beyond doubt by now. On the other hand, four--point functions of 
chiral primaries are not entirely constrained by symmetries or non--renormalization theorems. As such, studying their structure on the supergravity side is a challenging task that will disclose valuable information on CFT's at strong coupling.

Although we do not provide all the details of our computation\footnote{Our Mathematica code is available upon request.}, we have sought to explain a few non--trivial steps in the remainder of this note. We are especially highlighting aspects of our computation that are distinctive of processes involving the exchange of gravitons in the bulk, and are not part of the standard tool--box of previous Mellin--space computations. 

The last section illustrates our general formula on a few particular examples. A Mathematica notebook containing our universal formula is up for grabs from the source of the arXiv version of the present publication. This formula is somewhat lengthy but simplifies tremendously whenever arbitrary values of the operator scaling dimensions are specified. It is also worth pointing out that deriving this formula is an entirely algebraic and automated task. Former approaches would traditionally have called on performing two complicated integrations over $\text{AdS}_{d+1}$ or solving recursively some differential equations~\cite{D'Hoker:1999ni}.

It would be interesting to try and reverse--engineer our expression to position--space as a combination of D--functions. D--functions~\cite{D'Hoker:1999pj} are tantamount to tree--level $n$--point contact interactions between scalars in $\text{AdS}$ (cf.~Figure 1). It is true more generally that for the trilinear couplings stemming from $\text{AdS}_5 \times S^5$ supergravity, any exchange diagram reduces to a finite sum of scalar quartic graphs~\cite{D'Hoker:1999ni}. However, a conjecture of Dolan, Nirschl and Osborn~\cite{Dolan:2006ec} asserts that it should be possible to cast as a very particular and constrained sum of D--functions the dynamical part of four--point functions of single--trace chiral primaries belonging to any $\left[ 0, p,0 \right]$ $SU(4)$. This proposal for a specific combination of D--functions proceeds from an observation concerning the contribution of long multiplets with twist less than $2\, p$. We are aware of  at least one proposal for proving this conjecture\footnote{We are grateful to Leonardo Rastelli for explanations on this undertaking.}, which relies on bootstrapping arguments. Our result would provide a cross--check and lend further credence to such an attempt, were the latter be one day brought about with success.

Last but not least, the computation herein might also be viewed as one step closer to computing the long--sought four--point function of the $\mathcal{N}=4$ sYM stress--tensor, a demanding goal on which we comment in Section 3.
\begin{figure}
\centerline{
\includegraphics[width=0.4\textwidth]{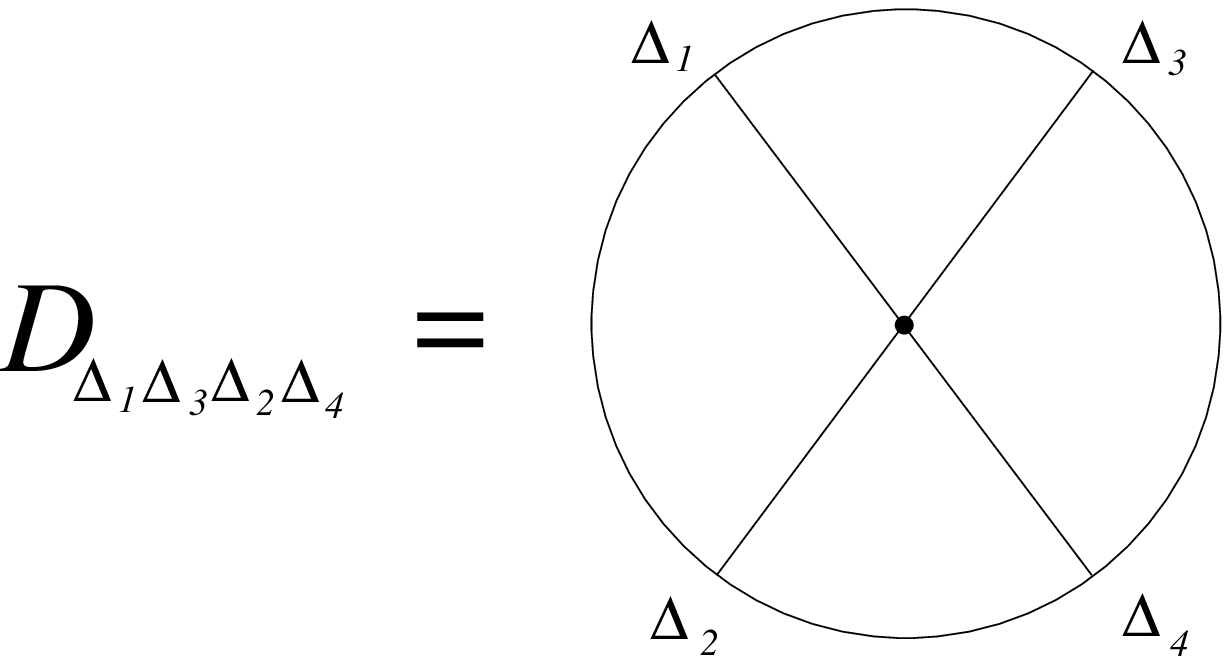}
}
\caption{\sl The D--function~\cite{D'Hoker:1999pj} for dual operators of respective conformal weight $\Delta_1$, $\Delta_2$, $\Delta_3$ and $\Delta_4$.
\label{fig:Dfct}}
\end{figure}

\newpage

\section{A review of the Mellin--transform of Witten diagrams}

Given a correlation function of $n$ scalar primary operators of conformal weights $\Delta_i$, $i=1,...,n$, living in a $d$--dimensional space, 
\be
\mathcal{A} \left( x_1, ..., x_n \right) \equiv \langle \mathcal{O}\left( x_1 \right) ...\, \mathcal{O}\left( x_n \right) \rangle \, ,
\ee
its Mellin--transform $\mathcal{M}\left( \delta_{ij} \right)$ is defined as follows:
\bea\label{Mellin transf}
&& \mathcal{A} \left( x_1, ..., x_n \right) = \frac{\mathcal{N}}{2\, \pi \, i}\, \int d\delta_{ij}\, \mathcal{M}\left( \delta_{ij}\right)\, \prod_{i \leq j}^{n} \Gamma\left[ \delta_{ij} \right]\, \left( x_{i} - x_{j} \right)^{-2\, \delta_{ij}} \, , \nonumber\\
&& \mathcal{N} \equiv \frac{\pi^{d/2}}{2}\, \Gamma\left[ \frac{\sum_{i=1}^{n} \Delta_i - d}{2} \right]\, \prod_{i=1}^{n} \frac{1}{2\, \pi^{d/2}\, \Gamma\left[ \Delta_i - \frac{d}{2} + 1 \right]} \, .
\eea
The parameters $\delta_{ij} \equiv \delta_{ji}$ are such that, for each index $i$, we have
\be
\sum_{j=1}^{n} \delta_{ij} = \Delta_{i} \, , \qquad \delta_{ii} = 0 \, .
\ee  
The integration contours for the $\frac{n\, \left( n-3 \right)}{2}$ independent integration variables above are chosen parallel to the imaginary axis and such that the real parts of the arguments of the gamma functions entering the integrand are positive.

In keeping with the idea of an effective conformal field theory for the low--lying spectrum of the dilation operator of an CFT~\cite{Heemskerk:2009pn, Heemskerk:2010ty, Fitzpatrick:2010zm, ElShowk:2011ag}, it is illuminating to think of the $\delta_{ij}$'s as related to a sort of Mandelstam invariants built out of the ``momentum" vectors $k_i$ for states whose invariant masses are actually the operator dimensions $\Delta_{i}$:
\be
- k_i^2 \equiv \Delta_{i} \, , \qquad \sum_{i=1}^{n} k_i \equiv 0 \, , \qquad \delta_{ij} \equiv k_i \cdot k_j \, 
\ee
and
\be
s_{i_1 ... i_{\ell} } \equiv - \left( \sum_{i_{m}=1}^{\ell} k_{i_{m}} \right)^2 \, ,
\ee
so that in particular
\be\label{Mandel2}
s_{ij} = - \left( k_i + k_j \right)^2 = \Delta_{i} + \Delta_{j} - 2\, \delta_{ij} \, .
\ee

This interpretation is especially apt since, as it turns out, the Mellin amplitude of a given Witten diagram is meromorphic with simple poles at $s_{ij}$. As such, it can be thought of as a natural basis for operator product expansions. Another catchword encapsulates a Mellin--transform as a kind of Fourier transform in the radial directions $\mid x_i - x_j \mid$. Those distances typically appear in the OPE of two operators inserted at $x_i$ and $x_j$. 

The Mellin--transform can be put to good use in a variety of situations, as illustrated a little while back~\cite{Paulos:2012nu} in the context of finding relations between some of the multi--loop integrals of interest to QCD--like theories. Here, our main focus revolves around computing Witten diagrams. According to the holographic dictionary, such scattering processes in the bulk of a weakly--curved asymptotically $\text{AdS}_{d+1}$ space translate into correlation functions for the operators of a dual strongly--coupled $CFT_{d}$. 

In this setting, it is especially helpful to combine the Mellin--transform with the so--called embedding formalism~\cite{Dirac:1936fq, Mack:1969rr, Penedones:2007ns, Cornalba:2009ax, Weinberg:2010fx}. This simply amounts to taking advantage of the fact that an $\text{AdS}_{d+1}$ space can be viewed as an hyperboloid embedded in an $d+2$--dimensional Minkowski space $\mathbb{M}^{d+2}$ with light--cone metric
\be\label{Mink metr}
ds^2 = -dX^{+}\, dX^{-} + \sum_{m,n=1}^{d}\, \delta_{m\, n} dX^{m}\, dX^{n} \equiv \eta_{M\, N}\, dX^{M}\, dX^{N} \, .
\ee
The boundary of $\text{AdS}_{d+1}$ corresponds to the set of null rays:
\be
\left\{ P \in \mathbb{M}^{d+2}\, :\, P^2 = 0, \, P \sim \lambda\, P, \, \lambda \in \mathbb{R} \right\} \, .
\ee
We can parametrize a vector $P$ describing the boundary of $AdS_{d+1}$ as
\be
P = \left( 1,\, x^2,\, x^{\mu} \right) \, ,
\ee
which makes contact with the more conventional description. In particular, note that
\be
P_{ij} \equiv -2\, P_i \cdot P_j = \left( x_i - x_j \right)^2 \, .
\ee
This quantity shows up in the definition of a generic Mellin--transform~\eqref{Mellin transf}.

The key advantage of the embedding formalism is that it linearizes the action of the conformal group.   We have implemented this property in our calculation of an universal formula for the stress--tensor contribution to a generic Green's function of four scalar primaries at strong coupling. This involves evaluating a Witten diagram connecting two pairs of massive or massless scalars via a graviton in the bulk of $\text{AdS}$. The two vertices involve covariant derivatives in $\text{AdS}$. Quite nicely, this translates into simple partial derivatives in the flat Minkowski embedding space.

Many more details can be found in~\cite{Paulos:2011ie}, whose conventions we are using.

\section{The graviton bulk--to--bulk propagator}

In position space, the physical part of the graviton bulk--to--bulk propagator between a point $z$ in $AdS_{d+1}$ and another point of coordinates $w$ is given by
\be\label{D'Hoker propagator}
G_{A\, B;\, A^{\prime}\, B^{\prime}}(z,w) = \left[ \partial_{A} \partial_{A^{\prime}}\, u\, \partial_{B} \partial_{B^{\prime}}\, u + \partial_{A} \partial_{B^{\prime}}\, u\, \partial_{B} \partial_{A^{\prime}}\, u \right]\, G(u) + g_{AB}\, g_{A^{\prime}B^{\prime}}\, H(u) \, .
\ee
With the following $\text{AdS}_{d+1}$ metric
\be
ds^2 = \frac{1}{z_0^2} \Big( dz_0^2 + \sum_{i=1}^{d} dz_i^2 \Big) \, ,
\ee 
the chordal distance $u$ is given by
\be
u \equiv \frac{\left( z - w \right)^2}{2\, z_0\, w_0} \, , \qquad (z-w)^2 = \delta_{CD} \left( z-w \right)_C\, \left( z-w \right)_D \, .
\ee 
The function $G(u)$ above denotes the massless scalar propagator in $\text{AdS}$,
\be
G(u) = \frac{\Gamma \left[\frac{d+1}{2}\right]}{d\, \left( 4\, \pi \right)^{\frac{d+1}{2}}}\, \left( 2\, u^{-1} \right)^{d}\, _2F_1\left[d, \frac{d+1}{2}; d+1; -2\, u^{-1} \right] \, .
\ee
The function $H(u)$ sitting next to the other tensor structure is expressed as~\cite{D'Hoker:1999jc, D'Hoker:1999pj}
\begin{align}
& H(u) = - \frac{1}{d -1}\, \left[ 2\, \left( 1 + u \right)^2\, G(u) + 2\, \left( d - 2 \right)\, p^{\prime}(u) \right] \, , \\
& p^{\prime}(u) = \frac{-2}{d\, \left( d - 1 \right)}\, \frac{\Gamma \left[\frac{d+1}{2}\right]}{\left( 4\, \pi \right)^{\frac{d+1}{2}}}\, \left( 2\, u^{-1} \right)^{d-1}\, _2F_1\left[d - 1, \frac{d+1}{2}; d+1; -2\, u^{-1} \right] \, . \label{p(u)}
\end{align}

There exists an expression for the massless scalar propagator $G(u)$ in the split and embedding formalism that is suitable for Mellin--space computations~\cite{Penedones:2010ue, Paulos:2011ie},
\begin{align}\label{split scalar propagator}
& G(u) = \int \frac{dc}{2\, \pi\, i}\, f_{0}(c)\, \int_{\partial \text{AdS}} dQ\, \int \tilde{d^2 s_{c}}\, e^{2\, s\, Q \cdot X + 2\, \bar{s}\, Q \cdot Y} \, , \nonumber\\
& \tilde{d^2 s_c} \equiv \frac{ds}{s}\, \frac{d\bar{s}}{\bar{s}}\, s^{h+c}\, \bar{s}^{h-c}\, , \qquad f_{0}(c) \equiv \frac{1}{2\, \pi^{2\, h}\, \left( h^2 - c^2 \right)}\, \frac{1}{\Gamma \left[ c \right]\, \Gamma \left[ -c \right]} \, ,
\end{align}
where we have introduced
\be
h \equiv \frac{d}{2} \, .
\ee

It would be natural to use a formula of this type for the graviton bulk--to--bulk propagator in order to compute the Witten diagram of present interest. A candidate for this propagator has been proposed in~\cite{Balitsky:2011tw}:
\bea\label{Balitsky propagator}
G^{A\, B}_{A^{\prime}\, B^{\prime}}(X;Y) &=& \int \frac{dc}{2\, \pi\, i}\, f_{0}(c)\, \left[ \left( h + 1 \right)^2 - c^2 \right]\, \int_{\partial \text{AdS}} dQ\, \int \frac{ds}{s}\, s^{h+c}\, \left( D_{h+c}^{M\, N\, A\, B\,}\, e^{2\, s\, Q \cdot X} \right)\, \times\nonumber\\ &\times& \, \mathcal{E}_{M\, N\, P\, Q}\, \int \frac{d\bar{s}}{\bar{s}}\, \bar{s}^{h-c}\, \left( D^{P\, Q}_{h-c\, \, \, A^{\prime}\, B^{\prime}}\, e^{2\, \bar{s}\, Q \cdot Y}\right)\, ,
\eea
where
\be
D_{\Delta}^{M_1\, M_2\, A_1\, A_2} \equiv \eta^{M_1\, A_1}\, \eta^{M_2\, A_2} + \frac{1}{\Delta}\, \left( \eta^{M_1\, A_1}\, P^{A_2}\, \partial_{M_2} + \left( 1 \leftrightarrow 2 \right)\, \right) + \frac{P^{A_1}\, P^{A_2}}{\Delta\, \left( \Delta + 1 \right)}\, \partial_{M_1}\partial_{M_2} \, 
\ee
and
\be
\mathcal{E}_{M\, N\, P\, Q} \equiv \frac{1}{2}\, \left( \eta_{M\, P}\, \eta_{N\, Q} + \eta_{M\, Q}\, \eta_{N\, P} \right) - \frac{1}{d}\, \eta_{M\, N}\, \eta_{P\, Q} \, ,
\ee
with the metric components $\eta_{M\, N}$ being the ones of a $d+2$--dimensional Minkowski space, as defined in equation~\eqref{Mink metr}.
This cannot be the right expression though. For instance, albeit it correctly reproduces the poles (and their residues) entering the known expression for the Mellin transform of the Witten diagram between four minimally--coupled scalars in $\text{AdS}_5$, it fails to yield the finite part accurately, namely the final piece $ - \frac{5}{4}\, \left(3\, s_{12} - 22\right)$ in equation~\eqref{mincoupled} below. The whole point is that~\eqref{Balitsky propagator} is traceless while the graviton propagator is not, actually. 

An alternative embedding--formalism and split--formalism expression for the graviton propagator appears in~\cite{Rizzo}. Unfortunately, it turns out that this proposal is riddled with its own shortcomings and does not reproduce~\eqref{mincoupled} either. A search for the correct formula is still on~\cite{wip} but has not proved fruitful so far. 

In view of those unexpected hindrances, any Mellin--transform of a Witten diagram involving the exchange of a graviton in the bulk of $\text{AdS}_{d+1}$ will have to make direct use of the position--space formula~\eqref{D'Hoker propagator}. This introduces some factor of $X \cdot Y$, $P_{i} \cdot X$ or $P_{j} \cdot Y$. Transforming those stray $X$'s and $Y$'s into an expression appropriate to a Mellin--space formulation requires some care, much more than if an out--of--the--box split formula for the graviton propagator were available. Still, it is an entirely straightforward task and once rules for doing so have been written, they can be recycled for the computation of any Witten diagram, which is thereby reduced to a simple and automated process. In particular, it is possible to combine our result with previous evaluations of other Witten diagrams. That would thus provide the missing ingredient to wrap up the supergravity computation of the planar four--point Green's function for scalars of arbitrary scaling dimension.

Also of much interest, it should be possible to compute the four--point function of the $\mathcal{N}=4$ sYM stress--tensor by that means~\cite{wip2}. The four--point Green's function of the stress--tensor is available for any two--dimensional CFT~\cite{Osborn:2012vt}; it is universal and depends only on the central charge $c$. Results have also been obtained fairly recently for three--dimensional CFT's with a supergravity dual~\cite{Raju:2012zs} using new recursion relations for AdS/CFT correlators~\cite{Raju:2012zr}, which are reminiscent of those of Risager for perturbative field theories~\cite{Risager:2005vk}. Clearly, an explicit $\text{AdS}_5$/$\text{CFT}_4$ calculation of this quantity is much sought--after\footnote{See~\cite{Costa:2011mg, Costa:2011dw, SimmonsDuffin:2012uy} for progress in this direction.}. Once again, the only valid reasons for not going ahead with such a computation mostly have to do with a bias against its presumed bulkiness and the expectation that a more concise and neater formula should be obtainable with even less effort if one were able to find a split expression for the bulk--to--bulk graviton propagator.

In the present work, such considerations of elegance have been temporarily swept aside. We are of the opinion that our universal formula for the 4--scalar graviton--exchange diagram is simple enough with regard to its universality, even though one might object that a snappier expression should be at hand using a, so far elusive, split formula for the graviton propagator similar to~\eqref{split scalar propagator} for the scalar propagator. If anything, it illustrates how powerful a Mellin--space formulation of AdS/CFT correlation functions really is, subsuming and generalizing extensively more than a decade--worth of efforts where 4--point functions of scalar operators have been computed only for quite restricted sets of scaling dimensions.

Even though we do not provide in this note all the details of the method that leads to our universal formula\footnote{A Mathematica notebook is available from the author.}, it is worth emphasizing that we are making use of the split formalism to the best of what is currently achievable. First of all, we are inserting into~\eqref{D'Hoker propagator} the split expression for the scalar propagator, namely equation~\eqref{split scalar propagator}. Furthermore, $H(u)$ itself can be cast into a much more convenient form. To do so, first of all, rewrite $H(u)$ by applying the second of Euler's hypergeometric transformations, i.e.
\be
_2F_1\left[a, b; c; z \right] = \frac{_2F_1\left[c-a, b; c; \frac{z}{z-1} \right]}{\left( 1 - z \right)^b} \, ,
\ee 
in order to express $p(u)^{\prime}$~\eqref{p(u)} in terms of $G(u)$ and its first--order derivative. We can then trade $_2F_1\left[ 2,\frac{d+1}{2},d+1,\frac{\frac{-2}{u}}{\frac{-2}{u}-1} \right]$, which is proportional to $p(u)^{\prime}$, for
\be
\frac{\frac{\frac{-2}{u}}{\frac{-2}{u}-1}}{\left(\frac{\frac{-2}{u}}{\frac{-2}{u}-1}\right)^{\prime}}\, \left(\frac{1}{\left(\frac{2}{u}\right)^{d}\, \left(1+\frac{2}{u}\right)^{-\frac{d+1}{2}}}\, G(u)\right)^{\prime} + \frac{1}{\left(\frac{2}{u}\right)^{d}\, \left(1+\frac{2}{u}\right)^{-\frac{d+1}{2}}}\, G(u) \, .
\ee
This results in
\be\label{Hu1}
H(u) = - 2\, u^2\, G(u) - \frac{2\, (2-d)\, u\, \left( 1-u^2 \right) G^{\prime}(u)}{(1-d)^2} \, .
\ee
On the other hand\footnote{The factors of $\mid X \cdot X \mid^{\frac{h+c}{2}}$ and $\mid Y\cdot Y \mid^{\frac{h-c}{2}}$ are there to enforce the conditions $X^2 = -R^2$ and $Y^2 = -R^2$ characterizing the embedding of an $AdS_{d+1}$ space of radius $R$ into a $d+2$--dimensional Minkowski space. As explained in Section 2, this condition is central to the Mellin--space formulation of AdS/CFT correlation functions.},
\be\label{Hu2}
G(u) = \int \frac{dc}{2\, \pi\, i}\, f_{0}(c)\, D_{c}(u) \, , \qquad D_{c}(u) \equiv \int_{\partial \text{AdS}} dQ\, \frac{\mid X \mid^{h+c}}{\left( - 2 \, X \cdot Q \right)^{h+c}}\frac{\mid Y \mid^{h-c}}{\left( - 2 Y \cdot Q \right)^{h-c}} \, .
\ee
For any function of the chordal distance $u$\footnote{From now on, we are going to define the chordal distance as $u = -X \cdot Y$. We have taken this into account when massaging the bulk--to--bulk propagator~\eqref{D'Hoker propagator} where $u$ is actually $u_D = -1 + u$ with our new convention.}, one can verify that
\be
\frac{\partial}{\partial u}f(u) = \frac{1}{1-u^2}\, \frac{\mid X \mid}{\mid Y \mid}\, Y^M\, \frac{\partial}{\partial X^M} f \, ,
\ee
so that in particular
\be\label{Hu3}
D_{c}(u)^{\prime} = \frac{h^2 - c^2}{2\, h\, \left( -1 + u^2 \right)}\, \Big[ D_{c+1}(u) + D_{c-1}(u) - 2\, u\, D_{c}(u) \Big] \, ,
\ee
using the fact that $D_{c}(u) = D_{-c}(u)$. All in all, combining~\eqref{Hu1},~\eqref{Hu2} and~\eqref{Hu3} brings in an expression for $H(u)$ similar to the one for the scalar propagator $G(u)$ in the split and embedding formalism, equation~\eqref{split scalar propagator}.

\newpage

\section{Evaluating the 4--scalar graviton--exchange Witten diagram}

The Mellin amplitude for the 4-scalar graviton--exchange Witten diagram is quite unwieldy when the masses of the four external scalars are left unspecified. Henceforth, to keep this note relatively short, in this section we refrain from writing down the generic expression. The reader is referred to the Mathematica notebook accompanying this paper for the details of the most general formula. Another Mathematica notebook is available upon request, where this end--product is derived step by step in a way that should also prove useful for computing other contributions to correlations functions at strong coupling.

For illustrative purposes, however, here we provide the Mellin amplitude for the Witten diagram corresponding to the exchange of a graviton in $\text{AdS}_{d+1}$ between scalars of masses $m_1^2  = d_1\, \left( d_1 - d \right)$  and $m_2^2 = d_2\, \left( d_2 - d \right)$, where the space--time dimension $d \equiv 2\, h$ of the boundary is further set to $d=4$. According to the AdS/CFT dictionary, the $d_i$'s correspond to the conformal weights of the dual operators.

Quite generally, our universal formula is expressed in terms of the following quantities, which are proportional to the Mellin--space formulation of scalar--exchange Witten diagrams:
\bea
\Upsilon[\Delta_1,\Delta_2;\delta; z] &\equiv& \frac{\Gamma \left[ \Delta_1+\frac{1}{2}\, \left(\delta -4 \right) \right]\, \Gamma \left[ \Delta_2 + \frac{1}{2}\, \left( \delta - 4 \right)\, \right]}{\Gamma [\delta -1] \left(z - \delta \right)} \times \nonumber\\ &\times& \, _3F_2 \left[ 1-\Delta_1 + \frac{\delta}{2} , 1-\Delta_2+\frac{\delta}{2} , \frac{\delta-z}{2} ; \delta - 1 , 1 + \frac{\delta - z}{2} ; 1\right] \, .
\eea 
Indeed, 
\bea
\Upsilon[\Delta_1,\Delta_2;\delta; z] &=& \int \frac{dc}{2\, \pi\, i}\, \frac{1}{\Gamma \left[\Delta_1 - \frac{z}{2}\right]\,\Gamma \left[\Delta_2 - \frac{z}{2}\right]}\, \Gamma \left[1+\frac{1}{2}(c-z)\right]\, \Gamma \left[1-\frac{1}{2}(c+z)\right]\, \times \nonumber\\ &\times&\, \frac{\Gamma \left[\Delta_1 - 1 + \frac{c}{2}\right]\, \Gamma \left[\Delta_1 - 1 -\frac{c}{2}\right]\, \Gamma \left[\Delta_2 - 1 + \frac{c}{2}\right]\, \Gamma \left[\Delta_2 -1 - \frac{c}{2}\right]}{4\, \left(c^2 - (\delta - 2)^2\right) \Gamma \left[c\right]\, \Gamma \left[-c\right]} \, .
\eea
Starting with our universal formula for the Mellin amplitude $\mathcal{M}\left( d_1, d_2, d_3, d_4; \gamma_{12}, s_{12};h\right)$ of the Witten diagram corresponding to the exchange of a graviton in $\text{AdS}_{2\, h + 1}$ between four bulk scalars of respective mass--squared $m_i^2 = d_i\, \left( d_i - 2\, h \right)$, we can particularize it to $h = 2$, $d_3 = d_1$ and $d_4 = d_2$. This leads to the following expression, spelt out in terms of the ``Mandelstam invariant'' $s_{12}$ (cf.~\eqref{Mandel2}) and the variable
\be
\gamma_{12} \equiv \frac{1}{2}\, \left( s_{13} - s_{23} \right) \, .
\ee
This particular case of our general expression for the Mellin amplitude of a 4--scalar graviton--exchange Witten diagram takes on the following form:
\bea
\mathcal{M}(d_1, d_2; \gamma_{12}, s_{12}) &=& \mathcal{P}_{(2,1)}\, \Upsilon[\frac{1}{2} \left(d_1+d_2\right)+2,\frac{1}{2} \left(d_1+d_2\right)+1; 4; s_{12}+2] \nonumber\\ 
&+& \mathcal{P}_{(1,1)}\, \Upsilon[\frac{1}{2} \left(d_1+d_2\right)+1,\frac{1}{2} \left(d_1+d_2\right)+1; 4; s_{12}+2] \nonumber\\ 
&+& \mathcal{P}_{(1,0)}\, \Upsilon[\frac{1}{2} \left(d_1+d_2\right)+1,\frac{1}{2} \left(d_1+d_2\right); 4; s_{12}+2] \nonumber\\ 
&+&\mathcal{P}_{(0,0)}\, \Upsilon[\frac{1}{2} \left(d_1+d_2\right),\frac{1}{2} \left(d_1+d_2\right); 4; s_{12}+2] \nonumber\\ 
&+&\mathcal{P}_{(0,-1)}\, \Upsilon[\frac{1}{2} \left(d_1+d_2\right),\frac{1}{2} \left(d_1+d_2\right)-1; 4; s_{12}+2] \nonumber\\ 
&+&\mathcal{P}_{(-1,-1)}\, \Upsilon[\frac{1}{2} \left(d_1+d_2\right)-1,\frac{1}{2} \left(d_1+d_2\right)-1; 4; s_{12}+2] \nonumber\\ 
&+& \mathcal{P}_{\frac{3}{2},\frac{3}{2}}\, \Upsilon[\frac{1}{2} \left(3+d_1+d_2\right),\frac{1}{2} \left(3+d_1+d_2\right); 3; s_{12}+1]\nonumber\\ 
&+& \mathcal{P}_{\frac{3}{2},\frac{1}{2}}\, \Upsilon[\frac{1}{2} \left(3+d_1+d_2\right),\frac{1}{2} \left(1+d_1+d_2\right); 3; s_{12}+1]
\nonumber\\ 
&+& \mathcal{P}_{\frac{1}{2},\frac{1}{2}}\, \Upsilon[\frac{1}{2} \left(1+d_1+d_2\right),\frac{1}{2} \left(1+d_1+d_2\right); 3; s_{12}+1]
\nonumber\\ 
&+& \mathcal{P}_{\frac{1}{2},-\frac{1}{2}}\, \Upsilon[\frac{1}{2} \left(1+d_1+d_2\right),\frac{1}{2} \left(-1+d_1+d_2\right); 3; s_{12}+1]
\nonumber\\ 
&+& \mathcal{P}_{-\frac{1}{2},-\frac{1}{2}}\, \Upsilon[\frac{1}{2} \left(-1+d_1+d_2\right),\frac{1}{2} \left(-1+d_1+d_2\right); 3; s_{12}+1] \, ,
\eea
where the
\be
\mathcal{P}_{(i,j)} = \mathcal{P}_{(i,j)}[d_1, d_2; \gamma_{12}, s_{12}] \, 
\ee
are given by
\bea
\mathcal{P}_{(2,1)} &=& -\frac{\pi^2}{16}\, \left(d_1 + d_2 - s_{12} \right)\, \big[ 8 + 4\, d_1^2 - 10\, d_2 + 2\, d_1\, \left( - 5 + 4\, d_2 - 2\, s_{12} \right) \nonumber\\ && + 5\, s_{12} + \left( - 2\, d_2 + s_{12} \right)^2 \big] \, ,
\eea
\bea
\mathcal{P}_{(1,1)} &=& \frac{\pi^2}{96}\, \Big\{ 7120 - 6192\, d_2 + 4\, \big( 9\, d_1^4 + d_1^3\, \left[ - 45 + 28\, d_2 \right] \nonumber\\ && + d_2^2\, \left( 422 + 9\, \left[ - 5 + d_2 \right]\, d_2 \right) + d_1^2\, \left( 422 + d_2\, \left[ - 99 + 38\, d_2 \right]\, \right) \nonumber\\ && + d_1\, \left[ - 1548 + d_2\, \left( 798 + d_2\, \left[ - 99 + 28\, d_2 \right]\, \right)\, \right]\, \big) \nonumber\\ 
&& + 3084\, s_{12} - s_{12}\, \big[ 48\, d_1^3 + d_1^2\, \left[ - 195 + 112\, d_2 \right] \nonumber\\ && + d_2\, \left( 1616 + 3\, d_2\, \left[ - 65 + 16\, d_2 \right]\, \right) + 2\, d_1\, \left( 808 + d_2\, \left[ - 159 + 56\, d_2 \right]\, \right)\, \big] \nonumber\\ &&
+ \big( 359 + 15\, d_1^2 + 3\, d_2\, \left[ - 14 + 5\, d_2 \right] + d_1\, \left[ - 42 + 22\, d_2 \right]\, \big)\, s_{12}^2 - 12\, \gamma _{12}^2 \Big\} \, , \nonumber\\
\eea
\begin{align}
& \mathcal{P}_{(1,0)} = - \frac{\pi ^2}{288\, \left( - 2 + d_1 + d_2 -s_{12}\right)}\, \Big\{ 54\, d_1^6 + 54\, d_2^6 + 9\, d_1^5\, \left( - 61 + 20\, d_2 - 10\, s_{12} \right) \nonumber\\ &
\, \, - 9\, d_2^5\, \left( 61 + 10\, s_{12} \right) + 9\, d_2^4\, \left( 907 + 90\, s_{12} + 4\, s_{12}^2 \right)
+ d_1^3\, \big( - 61404 + 2\, d_2\, \left( 12406 + d_2\, \left[ - 761 + 140\, d_2 \right]\, \right) \nonumber\\ &
\, \, - 4\, \left( 2694 + d_2\, \left[ - 447 + 89\, d_2 \right]\, \right)\, s_{12} + 3\, \left[ - 93 + 28\, d_2 \right]\, s_{12}^2 \big) - 3\, d_2^3\, \left( 20468 + s_{12}\, \left[ 3592 + 93\, s_{12} \right]\, \right) \nonumber\\ 
& \, \, - 8\, d_2\, \left( 88220 + s_{12}\, \left[ 26578 + 1713\, s_{12} \right]\, \right) + 4\, \left( 199224 + s_{12}\, \left[ 77546 + 7577\, s_{12}\, \right]\, \right) \nonumber\\
& \, \, + d_1^4\, \big( 266\, d_2^2 - 3\, d_2\, \left[ 451 + 86\, s_{12} \right] + 9\, \left[ 907 + 90\, s_{12} + 4\, s_{12}^2 \right] \big) + 2\, d_2^2\, \big( 2\, \left( 67715 + 3\, s_{12}\, \left[ 5353 + 272\, s_{12} \right]\, \right) \nonumber\\ &
\, \, - 9\, \left[ - 2 + s_{12} \right]\, \gamma_{12} \big) + d_1^2\, \big( 266\, d_2^4 - 2\, d_2^3\, \left[ 761 + 178\, s_{12} \right] + 2\, d_2^2\, \left( 16817 + 2\, s_{12}\, \left[ 515 + 26\, s_{12} \right]\, \right) \nonumber\\ 
& \, \, - 5\, d_2\, \left( 31700 + s_{12}\, \left[ 5176 + 105\, s_{12} \right]\, \right) + 6\, s_{12}\, \left( 10706 + 544\, s_{12} - 3\, \gamma_{12} \right) + 4\, \left[ 67715 + 9\, \gamma_{12} \right]\, \big) \nonumber\\
& \, \, + d_1\, \big[ 180\, d_2^5 - 3\, d_2^4\, \left( 451 + 86\, s_{12} \right) + 4\, d_2^3\, \left( 6203 + 3\, s_{12}\, \left[ 149 + 7\, s_{12} \right]\, \right) \nonumber\\
& \, \, - 5\, d_2^2\, \left( 31700 + s_{12}\, \left[ 5176 + 105\, s_{12} \right]\, \right) - 8\, \left( 88220 + s_{12}\, \left[ 26578 + 1713\, s_{12} \right]\, \right) + 4\, d_2\, \big( 9\, \left[ - 2 + s_{12} \right]\, \gamma _{12} \nonumber\\ 
& \, \, + 2\, \left( 63515 + s_{12}\, \left[ 14569 + 685\, s_{12} \right]\, \right) \big)\, \big]\, \Big\} \, , \nonumber\\
\end{align}
\begin{align}
& \mathcal{P}_{(0,0)} = \frac{\pi^2}{288\, \left( - 2 + d_1 + d_2 - s_{12} \right)^2}\, \Big\{ 9\, d_1^8 + 9\, d_2^8 - 18\, d_1^7\, \left( 7 + s_{12} \right) - 18\, d_2^7\, \left( 7 + s_{12} \right) \nonumber\\ &
\, \, + 9\, d_2^6\, \left( 407 + s_{12}\, \left[ 26 + s_{12} \right]\, \right) + 96\, \left( 117888 + s_{12}\, \left[ 56986 + 5901\, s_{12} \right]\, \right) \nonumber\\ &
\, \, - 8\, d_2\, \left( 1822776 + s_{12}\, \left[ 683202 + 55205\, s_{12} \right]\, \right) + 3\, d_1^6\, \left[ - 44\, d_2^2 - 2\, d_2\, \left( - 47 + s_{12} \right)+ 3\, \left( 407 + s_{12}\, \left[ 26 + s_{12} \right]\, \right)\, \right] \nonumber\\ 
& \, \, - 3\, d_2^5\, \left( 16468 + s_{12}\, \left( 2053 + 36\, s_{12} \right) + 6\, \gamma_{12} \right) - d_1^5\, \big( 49404 + 6\, d_2\, \left( - 1659 + d_2\, \left[ - 425 + 64\, d_2 \right]\, \right) + 6159\, s_{12} \nonumber\\ 
& \, \, + 2\, \left( 138 - 79\, d_2 \right)\, d_2\, s_{12} - 6\, \left( - 18 + d_2 \right)\, s_{12}^2 + 18\, \gamma_{12} \big) + 4\, d_2^2\, \big( 1947468 + 561902\, s_{12} + 34925\, s_{12}^2 \nonumber\\ & \, \, + 36\, \left( 2 + s_{12} \right)\, \gamma_{12} \big) + 6\, d_2^4\, \left( 68524 + 11971\, s_{12} + 416\, s_{12}^2 + 3\, \left( 8 + s_{12} \right)\, \gamma_{12} \right) \nonumber\\ &
\, \, - 4\, d_2^3\, \left( 568704 + 90\, \gamma_{12} + s_{12}\, \left[ 127018 + 6081\, s_{12} + 27\, \gamma_{12} \right]\, \right) \nonumber\\
& \, \, + d_1^2\, \big( - 132\, d_2^6 + 2\, d_2^5\, \left( 1275 + 79\, s_{12} \right) + d_2^4\, \left( 9969 - s_{12}\, \left[ 2566 + 37\, s_{12} \right]\, \right) + 2\, d_2^3\, \big( s_{12}\, \left( - 9411 + 238\, s_{12} \right) \nonumber\\ &
\, \, + 18\, \left[ - 7130 + \gamma_{12} \right]\, \big) - 4\, d_2\, \left( 1545696 + s_{12}\, \left( 324414 + 14107\, s_{12} - 27\, \gamma_{12} \right) - 90\, \gamma_{12} \right) \nonumber\\
& \, \, + 4\, \left( 1947468 + 561902\, s_{12} + 34925\, s_{12}^2 + 36\, \left( 2 + s_{12} \right)\, \gamma_{12} \right) + 4\, d_2^2\, \big[ 459036 + 67615\, s_{12} + 1690\, s_{12}^2 \nonumber\\ &
\, \, - 9\, \left( 8 + s_{12} \right)\, \gamma_{12} \big]\, \big) + d_1^4\, \big[ - 522\, d_2^4 + 6 d_2^3\, \left( 893 + 63\, s_{12} \right) + d_2^2\, \left( 9969 - s_{12}\, \left[ 2566 + 37\, s_{12} \right]\, \right) \nonumber\\
& \, \, + 6\, \left( 68524 + 11971\, s_{12} + 416\, s_{12}^2 + 3\, \left( 8 + s_{12} \right)\, \gamma_{12} \right) + d_2\, \left( s_{12}\, \left( - 15787 + 24\, s_{12} \right) - 18\, \left[ 9198 + \gamma_{12} \right]\, \right)\, \big] \nonumber\\
& \, \, + d_1\, \big[ - 6\, d_2^6\, \left( - 47 + s_{12} \right) + 6\, d_2^5\, \big( 1659 +\left( - 46 + s_{12} \right)\, s_{12} \big) + 8\, d_2^3\, \left( 165792 + s_{12}\, \left[ 25705 + 721\, s_{12} \right]\, \right) \nonumber\\
& \, \, - 8\, \left( 1822776 + s_{12}\, \left[ 683202 + 55205\, s_{12} \right]\, \right) - 4\, d_2^2\, \left( 1545696 + s_{12}\, \left( 324414 + 14107\, s_{12} - 27\, \gamma_{12} \right) - 90\, \gamma_{12} \right) \nonumber\\
& \, \, + 8\, d_2\, \left( 1882956 + 531278\, s_{12} + 31901\, s_{12}^2 - 36\, \left( 2 + s_{12} \right)\, \gamma _{12} \right) \nonumber\\ &
\, \, + d_2^4\, \left( s_{12}\, \left( - 15787 + 24\, s_{12} \right) - 18\, \left[ 9198 + \gamma_{12} \right]\, \right)\, \big] + 2\, d_1^3\, \big[ - 192\, d_2^5 + 3\, d_2^4\, \left( 893 + 63\, s_{12} \right) \nonumber\\ &
\, \, + d_2^3\, \left( 3678 - 2\, s_{12}\, \left[ 1028 + 17\, s_{12} \right]\, \right) + 4\, d_2\, \left( 165792 + s_{12}\, \left[ 25705 + 721\, s_{12} \right]\, \right) + d_2^2\, \big( s_{12}\, \left( - 9411 + 238\, s_{12} \right) \nonumber\\ 
& \, \, + 18\, \left[ - 7130 + \gamma_{12} \right]\, \big) - 2\, \big( 568704 + 90\, \gamma_{12} + s_{12}\, \left[ 127018 + 6081\, s_{12} + 27\, \gamma_{12} \right]\, \big)\,\big]\, \Big\} \, , \nonumber\\
\end{align}
\begin{align}
& \mathcal{P}_{(0,-1)} = \frac{\pi^2}{72}\, \frac{- 3 + d_1 + d_2}{\big( - 4 + d_1 + d_2 - s_{12} \big)\, \big( - 2 + d_1 + d_2 - s_{12} \big)^2}\, \Big\{ 3\, d_1^8\, d_2 \nonumber\\ &
+ 3 d_1^7\, \big( - 40 + d_2\, \big[ - 25 + 5\, d_2 - 2\, s_{12} \big]\, \big) + d_1^5\, \big( - 41952 + 3892\, d_2 + 3068\, d_2^2 - 665\, d_2^3 \nonumber\\
& + 45\, d_2^4 - 2\, \big( 2640 + d_2\, \big[ - 43 + 5\, d_2 \big]\, \big[ - 7 + 5\, d_2 \big]\, \big)\, s_{12} + \big( - 120 + d_2\, \big[ - 57 + 11\, d_2 \big]\, \big)\, s_{12}^2 \big) \nonumber\\
& + d_1^4\, \big( d_2\, \big( - 121312 + d_2\, \big[ - 7988 + 9\, d_2\, \big( 676 + 5\, \big( - 18 + d_2 \big)\, d_2 \big)\, \big]\, \big) \nonumber\\
& - 2\, d_2\, \big( 3698 + d_2\, \big[ 1783 + 30\, \big( - 14 + d_2 \big)\, d_2 \big]\, \big)\, s_{12} + \big[ 2280 + d_2\, \big( 170 + 3\, d_2\, \big[ - 59 + 6\, d_2 \big]\, \big)\, \big]\, s_{12}^2 \nonumber\\ & 
+ 208\, \big[ 1816 + 313\, s_{12} \big]\, \big) + d_1^2\, \big( d_2\, \big( - 5773568 + d_2\, \big[ 1648256 + d_2\, \big( - 160688 \nonumber\\ 
& + d_2\, \big[ - 7988 + d_2\, \big( 3068 + 5\, d_2\, \big[ - 67 + 3\, d_2 \big]\, \big)\, \big]\, \big)\, \big]\,\big) + 2\, d_2\, \big( - 630320 + d_2\, \big[ 101316 + d_2\, \big( 694 \nonumber\\ &
+ d_2\, \big[ - 1783 + \big( 250 - 13\, d_2 \big)\, d_2 \big]\, \big)\, \big]\, \big)\, s_{12} + \big[ 150816 + d_2\, \big( - 50992 + d_2\, \big[ 2428 + d_2\, \big( 866 \nonumber\\ &
+ d_2\, \big[ - 177 + 11\, d_2 \big]\, \big)\, \big]\, \big)\, \big]\, s_{12}^2 + 960\, \big[ 7302 + 2347\, s_{12} \big]\, \big) + 8\, \big( d_2\, \big[ - 1563136 + d_2\, \big( 876240 \nonumber\\ 
& + d_2\, \big[ - 264416 + d_2\, \big( 47216 - 3\, d_2\, \big[ 1748 + 5\, \big( - 25 + d_2 \big)\, d_2 \big]\, \big)\, \big]\, \big)\, \big] \nonumber\\ &
+ 2\, d_2\, \big( - 335896 + d_2\, \big[ 140820 + d_2\, \big( - 31294 + d_2\, \big[ 4069 + 15\, \big( - 22 + d_2 \big)\, d_2 \big]\, \big)\, \big]\, \big)\, s_{12} \nonumber\\
& + \big( 81536 + d_2\, \big[ - 61736 + d_2\, \big( 18852 + d_2\, \big[ - 3014 - 15\, \big( - 19 + d_2 \big)\, d_2 \big]\, \big)\, \big]\, \big)\, s_{12}^2 + 256\, \big[ 4483 + 2556\, s_{12} \big]\, \big) \nonumber\\ &
+ d_1^3\, \big( d_2\, \big[ 1199488 + d_2\, \big( - 160688 + d_2\, \big[ - 17760 + d_2\, \big( 6084 + d_2\, \big[ - 665 + 33\, d_2 \big]\, \big)\, \big]\, \big)\, \big] \nonumber\\
& + 2\, d_2\, \big( 82472 + d_2\, \big[ 694 + d_2\, \big( - 2724 + 5\, \big( 84 - 5\, d_2 \big)\, d_2 \big)\,\big]\, \big)\, s_{12} + 2\, \big( - 12056 + d_2\, \big[ 1644 \nonumber\\
& + d_2\, \big( 433 + 3\, d_2\, \big[ - 40 + 3\, d_2 \big]\, \big)\, \big]\, \big)\, s_{12}^2 - 32\, \big[ 66104 + 15647\, s_{12} \big]\, \big)+ d_1\, \big( d_2\, \big( 13594880 \nonumber\\
& + d_2\, \big[ - 5773568 + d_2\, \big( 1199488 + d_2\, \big[ - 121312 + d_2\, \big( 3892 + 3\, d_2\, \big[ 152 + \big( - 25 + d_2 \big)\, d_2 \big]\, \big)\, \big]\,\big)\, \big]\, \big) \nonumber\\
& - 2\, d_2\, \big[ - 2132160 + d_2\, \big( 630320 + d_2\, \big[ - 82472 + d_2\, \big( 3698 + d_2\, \big[ 301 + 3\, \big( - 22 + d_2 \big)\, d_2 \big]\, \big)\, \big]\, \big)\, \big]\, s_{12} \nonumber\\
& + \big[ - 493888 + d_2\, \big( 271552 + d_2\, \big[ - 50992 + d_2\, \big( 3288 + d_2\, \big[ 170 + 3\, \big( - 19 + d_2 \big)\, d_2 \big]\, \big)\, \big]\, \big)\, \big]\, s_{12}^2 \nonumber\\
& - 128\, \big[ 97696 + 41987\, s_{12} \big]\, \big) + d_1^6\, \big[ 33\, d_2^3 + 120\, \big( 25 + 2\, s_{12} \big) - d_2^2\, \big( 335 + 26\, s_{12} \big) \nonumber\\ &
+ 3\, d_2\, \big( 152 + s_{12}\, \big[ 44 + s_{12} \big]\, \big)\, \big]\, \Big\} \, ,
\end{align}
\bea
\mathcal{P}_{(-1,-1)} &=& \frac{\pi^2}{72}\, \frac{\left( - 5 + d_1 + d_2 \right)\, \left( - 4 + d_1 + d_2 \right)^2\, \left( - 3 + d_1 + d_2 \right)}{\left( - 4 + d_1 + d_2 - s_{12} \right)\, \left( - 2 + d_1 + d_2 - s_{12} \right)} \times \nonumber\\ 
&\times& \Big\{ d_1^4\, d_2^2 + 64\, \left( 516 + 25\, \left[ - 8 + d_2 \right]\, d_2 \right) + 2\, d_1^3\, d_2\, \left( - 40 + \left[ - 4 + d_2 \right]\, d_2 \right) \nonumber\\ 
&& - 16\, d_1\, \left[ 800 + d_2\, \left( - 104 + 5\, \left[ - 8 + d_2 \right]\, d_2 \right)\, \right] \nonumber\\ 
&& + d_1^2\, \left( 1600 + d_2\, \left[ 640 + d_2\, \left( - 148 + \left[ - 8 + d_2 \right]\, d_2 \right)\, \right]\, \right)\, \Big\} \, ,
\eea
\bea
\mathcal{P}_{(\frac{3}{2},\frac{3}{2})} & = & - \frac{\pi^2}{16} \left( d_1 + d_2 - s_{12} \right)^2\, \left( 2\, \left[ 1 - d_1 - d_2 \right] + s_{12} \right) \, ,
\eea
\bea
\mathcal{P}_{\frac{3}{2},\frac{1}{2}} &=& - \frac{\pi^2}{96}\, \left( d_1 + d_2 - s_{12} \right)^2\, \Big\{ 4\, \big( - 329 + 3\, d_1^3 + d_1^2\, \left( - 9 + 5\, d_2 \right) \nonumber\\ &+& 
d_2\, \left( 175 + 3\, \left( - 3 + d_2 \right)\, d_2 \right) + 5\, d_1\, \left[ 35 + \left( - 2 + d_2 \right)\, d_2 \right]\, \big) \nonumber\\ &-& \left( 350 + 9\, d_1^2 + 3\, d_2\, \left( - 8 + 3\, d_2 \right) + 2\, d_1\, \left( - 12 + 5\, d_2 \right)\, \right)\, s_{12} \Big\} \, , \nonumber\\
\eea
\begin{align}
& \mathcal{P}_{\frac{1}{2},\frac{1}{2}} = \frac{\pi^2}{288}\, \Big\{ 9\, d_1^5 + d_1^3\, \left(2046+2\, \left(54 - 19\, d_2 \right)\, d_2 + 45\, s_{12} \right) - 4\, \left( 25920 + 5867\, s_{12} \right) \nonumber\\ &
- 3\, d_1^4\, \left( d_2 + 3\, \left[ 5+s_{12} \right]\, \right) + d_1^2\, \left( - 54\, \left( 217 + 29 s_{12} \right) + d_2\, \left[ 4138 - 33\, s_{12} + 2\, d_2\, \left( 129 - 19\, d_2 + 5\, s_{12} \right)\, \right]\, \right) \nonumber\\ &
+ d_2 \left( 54508 + 6918\, s_{12} + 3\, d_2\, \left[ d_2\, \left( 682 + 3\, d_2\, \left( - 5 + d_2 - s_{12} \right) + 15\, s_{12} \right) - 18\, \left( 217 + 29\, s_{12} \right)\, \right]\, \right) \nonumber\\ &
+ d_1\, \left( 54508 + 6918\, s_{12} - d_2\, \left[ d_2\, \left( - 4138 + 3\, \left( - 36 + d_2 \right)\, d_2 + 33\, s_{12} \right) + 4\, \left( 4731 + 557\, s_{12} \right)\, \right]\, \right) \Big\} \, ,
\end{align}
\begin{align}
& \mathcal{P}_{\frac{1}{2},-\frac{1}{2}} = \frac{\pi^2}{72}\, \frac{\left(40-d_1\, d_2\right)}{\left(2-d_1-d_2+s_{12}\right)}\, \Big\{ 3\, d_1^5 - 64\, \left( 93 + 29\, s_{12} \right) + d_1^3\, \left( 406 + 3\, d_2 \left( - 44 + 6\, d_2 - 3\, s_{12} \right) + 36\, s_{12} \right) \nonumber\\ &
+ d_1^4\, \left( 11\, d_2 - 3\, \left[ 14 + s_{12} \right]\, \right) + 2\, d_1^2\, \left( - 21\, \left( 52 + 7\, s_{12} \right) + d_2\, \left[ 551 + 9\, d_2^2 + 44\, s_{12} - 6\, d_2\, \left( 15 + s_{12} \right)\, \right]\, \right) \nonumber\\ &
+ d_2\, \left( 5768 + 1196\, s_{12} + d_2\, \left( - 42\, \left( 52 + 7\, s_{12} \right) + d_2\, \left( 406 + 3\, d_2\, \left( - 14 + d_2 - s_{12} \right) + 36\, s_{12} \right)\, \right)\, \right) \nonumber\\
& + d_1\, \left( 5768 + 1196\, s_{12} + d_2\, \left( - 4236 - 550\, s_{12} + d_2\, \left( 1102 + d_2\, \left( - 132 + 11\, d_2 - 9\, s_{12} \right) + 88\, s_{12} \right)\right)\right) \Big\} \, 
\end{align}
and at long last
\bea
\mathcal{P}_{-\frac{1}{2},-\frac{1}{2}} &=& \frac{\pi^2}{72}\, \left(-4+d_1+d_2\right)\, \left(-3+d_1+d_2\right)\, \left(-40+d_1 d_2\right)^2 \times \nonumber\\ &\times& \frac{\left(14+d_1^2+2\, d_1 \left[-3+d_2\right]+\left[-6+d_2\right]\, d_2 \right)}{\left( -2+d_1+d_2 - s_{12} \right)}\, .
\eea

In particular, when all the $d_i$'s are set to $d_i = \Delta = 4$, we recover\footnote{See the affixed Mathematica file.} up to an overall normalization factor the known expression for the graviton--exchange between minimally--coupled scalars in $\text{AdS}_5$:
\be\label{mincoupled}
\mathcal{M}(\gamma_{12}, s_{12}) = \frac{6\, \gamma _{12}^2 + 2}{s_{12}-2} + \frac{\gamma _{12}^2 - 1}{s_{12} - 6} + \frac{8\, \gamma _{12}^2}{s_{12} - 4} - \frac{5}{4}\, \left(3\, s_{12} - 22\right) \, .
\ee
This is in perfect agreement with equation (8) from reference~\cite{Penedones:2010ue} ! It is a non--trivial test on the validity of our universal formula. Indeed, in~\cite{Penedones:2010ue} the Mellin amplitude~\eqref{mincoupled} is obtained by directly translating to Mellin--space a result first derived in position--space by entirely different means and expressed in terms of D--functions by D'Hoker and collaborators~\cite{D'Hoker:1999pj}.

Our universal formula --- to be found in the Mathematica notebook enclosed in the source for the arXiv version of this paper --- can be used to evaluate in no time the Witten diagram describing the exchange of a graviton between scalars of arbitrary masses in $\text{AdS}$. From a dual perspective, this is the stress--tensor contribution to the four--point function of scalar operators at strong coupling, whatever their conformal weights and the dimension of the space in which they are defined. For instance, the graviton--exchange between minimally--coupled scalars, now in $\text{AdS}_7$, takes on the following form:
\be
\frac{360\, \pi\, \left(4 +5\, \gamma _{12}^2\right)}{s_{12}-4} + \frac{810\, \pi\, \left(3 +5\, \gamma _{12}^2\right)}{s_{12}-6} + \frac{1620\, \pi\,  \gamma _{12}^2}{s_{12} - 8} - \frac{90\, \pi\, \left( 1 -  \gamma _{12}^2\right)}{s_{12}-10} + 1890\, \pi\, \left(10 - s_{12}\right) \, .
\ee
As another example among endless possible combinations, here is the $\text{AdS}_5$ graviton--exchange contribution to the CFT--dual Green's function involving external scalar operators of respective dimensions $d_1 = 3$, $d_2 = 4$, $d_3 = 5$ and $d_4 = 7$:
\begin{align}
& \frac{2023\, \sqrt{\pi}+ 2250\, \sqrt{\pi}\, \gamma_{12} + 6300\, \sqrt{\pi}\, \gamma_{12}^2}{128\, \left(s_{12} - 2\right)} + \frac{17251\, \sqrt{\pi} + 1950\, \sqrt{\pi}\, \gamma_{12} + 25200\, \sqrt{\pi}\, \gamma_{12}^2}{256\, \left(s_{12} - 4\right)} \nonumber\\ &
+ \frac{456889\, \sqrt{\pi} - 3825\, \sqrt{\pi}\, \gamma_{12} + 18900\, \sqrt{\pi}\, \gamma_{12}^2}{1024\, \left(s_{12} - 6\right)} + \frac{- 356779\, \sqrt{\pi} + 2430\, \sqrt{\pi}\, \gamma_{12} - 5040\, \sqrt{\pi}\, \gamma_{12}^2}{4096\, \left(s_{12} - 8\right)} \nonumber\\
& + \frac{3\, \left(59023\, \sqrt{\pi} - 320\, \sqrt{\pi}\, \gamma_{12} + 420\, \sqrt{\pi}\, \gamma_{12}^2\right)}{16384\, \left(s_{12} - 10\right)} + \frac{225225\, \sqrt{\pi}\, \left(44-3\, s_{12}\right)}{16384}\, .
\end{align}

\vskip 0.7cm
 \noindent {\bf Acknowledgements}:\\
 \noindent I would like to thank Balt van Rees for discussions, and especially Miguel Paulos for helpful collaboration at the beginning of this project and on an ongoing research endeavour. I am grateful to Miguel Paulos and to Leonardo Rastelli for comments on the manuscript. Funding by the Research Foundation, Stony Brook University is appreciated. I have also benefited in part from an ERC Starting Independent Researcher Grant 240210 --- String--QCD--BH.



\end{document}